# The Verifications of No Bright and Dark Fringes In a Standing-wave

JiWu Chen[1,*], Wen Wei[2]

[1]College of Metrology & Measurement Engineering, China Jiliang University, Hangzhou, 310018, China
[2]College of Modern Science & Technology, China Jiliang University, Hangzhou, 310018, China
*Corresponding author: chen_jiwu@aliyun.com

**Abstract:** After Wiener's experiment, it had been widely accepted that only electric-field was the fundamental-factor for optical interference and believed that there were alternating bright and dark fringes in a light standing-wave field which was formed by a pair of parallel light-beams traveling in mutually opposite directions. While being supplied to the bright fringes, the light-energy from the beams must pass across the dark fringes or across the electric-field nodes, so that the electric-field would not keep zero at the nodes. However, it was inconsistent with the characteristic of electric-field nodes at which the electric-field was zero anytime. Therefore, this naturally led to a presumption that there were no bright and dark interference-fringes in a standing-wave field, which was consistent with zero of the electromagnetic energy-flux-density in a standing-wave field. In this paper, it was fully verified from several aspects, including an ideal experiment, comparing a standing-wave with an LC circuit, a group of experiments and theoretical deductions of formulas. An inclination factor was introduced into the previous common formula of two-beams-interference to rectify it back to the original meaning of its initial definition by energy-flux-density $\vec{E} \times \vec{H}$ rather than electric-field energy-density $|E|^2$. Given the angle between the two-beams to be 180° as a validation, it was verified by the corrected formula that the light-intensities of the bright interference-fringes in a standing-wave were equal to zero. i.e., no interference-fringes existed in a standing-wave field. Thus, the photographic emulsion was blackened by electromagnetic energy-flux density. Instead of electric-field energy-density alone from Wiener's conclusions, interference-energy-flux density was the fundamental-factor to form interference-fringes. The magnetic-field vector acts the same role as the electric-field vector on light interacting with substance.

## 1 The Query Against Wiener's Experiment

It has been generally believed that only electric-field of light blackened the photographic emulsion since 1890 when Wiener experimentally demonstrated the existence of optical standing wave and then concluded that the photochemical action was directly related to the electric-field and not to the magnetic-field in light.[1] Let's review Wiener's experiment referring to Fig.1.1.

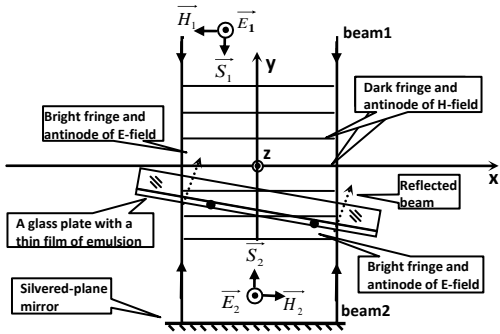

Fig.1.1. Wiener's experiment setup

While illuminating normally on a silvered-plane mirror, a parallel light beam1 interfered with the reflected beam2 to form a field of light standing-wave. A glass plate coated on one of its surfaces with a thin film of transparent photographic emulsion in thickness of less than 1/20 wavelength was placed in front of and inclined to the mirror at a small angle to detect the standing-wave. After development, equidistant parallel blackened fringes appeared on the emulsion. The blackened maxima corresponded to the intersection between the emulsion film and the antinodes of the electric-field, according to the further his experiments[1].

Several similar experiments were demonstrated subsequently [2-4]. In recent years, relative experiments were moreover accomplished [5-8].

In fact, the interference fringes appeared in Wiener's experiment were a kind of equal thickness fringes which were made by the two beams reflected from the film and the silvered-plane mirror respectively as shown in Fig.1.1. In other words, the fringes of Wiener's experiment were not formed by standing-waves.

As the reflectivity of the emulsion film was quite small relative to that of the plane mirror in Wiener's experiment, it was likely to make people mistake the fringe contrast as a small value and omit their equal thickness fringes.

The photos of typical equal-thickness fringes were presented in Fig.1.2, when a parallel light-beam illuminated an optical flat and a plane mirror as shown in Fig.1.3. Given the reflectivities on the optical-flat surface as 0.04 and the plane mirror as 1.00, the fringe contrast could reach 40% ( $V = \dfrac{2\sqrt{I_1 I_2}}{I_1 + I_2} = \dfrac{2\sqrt{0.04 \times 1}}{0.04 + 1} \approx 40\%$ ).

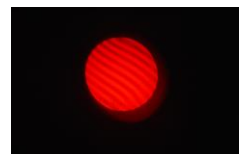  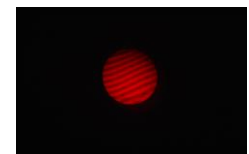

(a) A photo with F3.2 and 1/30 s    (b) A photo with F3.2 and 1/30 s

Fig.1.2. The equal thickness fringes between an optical flat and a plane mirror

The interference pattern was only determined by the angle between the plane mirror and the standard surface of the optical-flat. It did not correlate with the incident angle of the parallel light-beam. When a parallel light-beam illuminated the plane mirror normally referring to Fig.1.3, the fringes similar to Fig.1.2 (a) and (b) would also appear on the surface of the optical-flat. If the optical-flat was replaced with a photographic plate, the same fringes as on the optical-flat would be recorded on the photographic emulsion of it. So, the equal thickness fringes must have clearly appeared in Wiener's experiment. If the equal-thickness fringes had been seriously paid attention to in Wiener's experiment, it should not have been believed that a group of standing-wave fringes had hidden under the clear fringes similar to Fig.1.2 (a) and (b).

The subsequently similar experiments repeated the error. They had mistaken the equal thickness fringes as standing-wave fringes. [2-8]

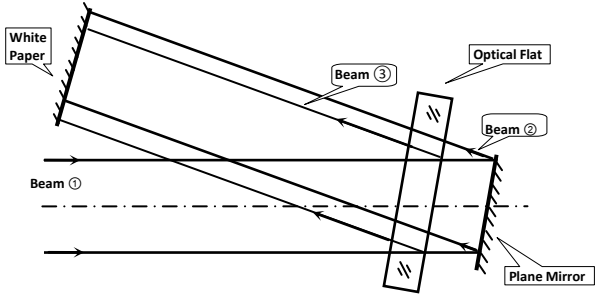

Fig.1.3. The setup for equal thickness fringes between an optical flat and a plane mirror

In this paper, it will be proved from several aspects that no alternating bright and dark interference fringes in a standing-wave field.

## 2 The Ideal Experiment against Interference Fringes in Standing-waves

If two parallel linearly polarized light-beams, $S_1$ and $S_2$, in xy-plane with angle $\theta$ between them and with the same wavelength $\lambda$ and with the electric-field vectors along z-axis were illuminating yz-plane, a group of alternating bright and dark interference-fringes would appear in the intersecting area, as shown in Fig.2.1. The 'bright' indicated that there must be electromagnetic energy-flux flowing along the bright fringes. The 'dark' indicated that there must be no any electromagnetic energy-flux flowing along the dark fringes.

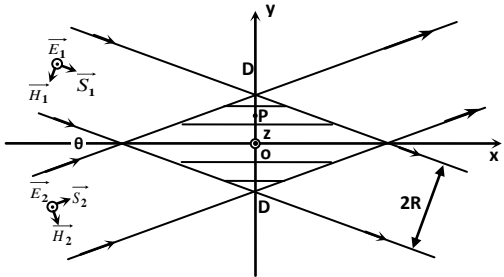

Fig.2.1. An interference field formed by two parallel light beams

If angle $\theta$ was getting bigger, the interference fringes would become denser. When the angle $\theta$ was equal to 180°, the spacing between interference fringes would reach to the minimum of $\lambda/2$. It was a standing-wave field described by the current optics[9].

In the standing-wave field which was formed by the opposite traveling light-beams $S_1$ and $S_2$, there must be nonzero light-energy or electromagnetic energy-flux-densities flowing along the 'bright' fringes to make them bright otherwise they would not be 'bright'. And there must be no light-energy going through the 'dark' fringes or the electromagnetic energy-flux-densities at the fringes must be zero otherwise the fringes would not be 'dark'. However, the flowing light-energy in the 'bright' fringes would be supplied by the two parallel light-beams $S_1$ and $S_2$ which passed across the 'dark' interference fringes or across the nodes of electric-field. In other words, the supplemental light-energy must pass across the 'dark' fringes, which made 'dark' fringes bright or made the electric-field at the nodes be nonzero.

So far, the ideal experiment has indicated that no alternating bright and dark interference-fringes exist in a standing-wave field. The indication was against the conclusions from the current light-interference principle, in which light interference-field was determined alone by electric-field energy density.

## 3 Equivalency between Standing-waves and LC Circuits

**[i]** *In a standing-wave field*

If a parallel light beam with polarization direction along x-axis illuminated a plane mirror, a standing-wave field was formed by the incident beam1 and reflected beam2 as shown in Fig.3.1.

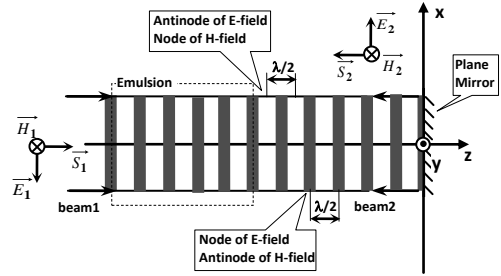

Fig.3.1. A standing-wave field formed by two parallel light beams

There were a series of alternating nodes and antinodes in the standing-wave field. The spacing between the antinodes of electric-field or magnetic-field was a half of wavelength $\lambda/2$. So there were some characteristics in the standing-wave field as follows[9]:

① Electric-field and magnetic-field were vibrating between their antinodes. As electric-field and magnetic-field were in quadrature, the average of electromagnetic energy-flux-density was equal to zero.

② As the amplitudes of electric-field or magnetic-field at their nodes were always equal to zero, the electromagnetic energy-flux never flowed through the nodes. It is a stationary wave, i.e. a standing wave.

③ Electric-field antinodes were just at the positions of magnetic-field nodes and magnetic-field antinodes were just at the positions of electric-field nodes. When the electric-field at an antinode was vibrating up to its maximum, the magnetic-field at the adjacent antinode was just reaching to zero at the same time. They were to alternate their roles. The magnetic-field reached just to its maximum when the electric-field reached to zero. In other words, the electromagnetic energy was vibrating between every pair of contiguous nodes in the way that electric-field energy and magnetic-field energy were converted into each other.

④ Standing-waves were considered only as a form of energy storages. No bright fringes existed because no energy-flux flowed through a standing-wave to make the fringes shine.

**[ii]** *An LC circuit equivalent to adjacent antinodes in a standing-wave*

As known in the electromagnetic theory, most of the electrical energy flowed into the load in electromagnetic waves through the space between the wires rather than through the inside of the wires, though the currents flowed through the inside of the wires as shown in Fig.3.2.

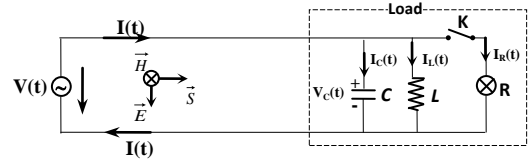

Fig.3.2. LC resonant circuit equivalent to a standing wave

Between the wires, the electric-field $E$ and magnetic-field $H$ were proportional to the voltage $V(t)$ and the current $I(t)$ respectively. So the electrical power $[V(t) \cdot I(t)]$ supplied for the load was proportional to the electromagnetic energy-flux-density or Poynting vector, $\vec{S}$ ($\vec{E} \times \vec{H}$).

If the frequency of the electrical source $V(t)$ was just the resonant frequency $(2\pi\sqrt{LC})^{-1}$ of an $LC$ parallel circuit, the circuit would go into a resonant state after the $LC$ circuit was powered on for a moment. Several characteristics similar to those in a standing-wave field were as follows.

① As the voltage $V_C(t)$ on C was equal to $V(t)$ in quadrature with the current $I_L(t)$, the electric-field in $C$ was in quadrature with the magnetic-field in $L$. In other words, the electric-field and the magnetic-field were vibrating between both $C$ and $L$.

In this way, it could be recognized in the $LC$ resonant circuit that the capacitance $C$ was corresponding to an electric-field antinode in a standing wave and the inductance $L$ was corresponding to the adjacent magnetic-field antinode.

② As $I_C(t)$ was in opposite phase with $I_L(t)$, $I(t)=[I_C(t)+I_L(t)]=0$. That is, no power $[I(t)\cdot V(t)]$ flowed into $LC$. So $LC$ was corresponding to a pair of two adjacent antinodes in a standing wave.

③ Electromagnetic energy was vibrating between $L$ and $C$ in the way that electric-field energy and magnetic-field energy were converted into each other alternately, which was corresponding to the electromagnetic energy vibrating between every pair of contiguous nodes in a standing-wave.

④ The $LC$ circuit was only a form of energy storages. No electrical energy was consuming because no energy-flux flowed into the $LC$ resonant circuit, which was corresponding to no bright fringes existing in a standing-wave.

Once the switch K in Fig.3.2 was turn on, the lamp would be lightened. In this case, there were two kinds of waves or currents with the same voltage $V(t)$ in the circuit. One was the sanding-wave corresponding to $V_C(t)$ and $I_L(t)$; the other one was a traveling electromagnetic wave corresponding to $V(t)$ and $I_R(t)$. It was $V(t)$ in phase with $I_R(t)$ that ensured the lamp lighting in Fig.3.2. The electrical power on the lamp was $V(t)\cdot I_R(t)$ where the voltage $V(t)$ and the current $I_R(t)$ played equivalent roles.

Similarly, there would be two kinds of waves in Fig.3.1 when a bulk of emulsion was placed in the standing wave. One was the standing wave as mentioned above; the other one was a traveling electromagnetic wave which blackened the emulsion. As traveling electromagnetic waves were not able to form any fringes, there were never any bright and dark fringes in the emulsion as well as in standing waves. As light-energy could be consumed only in the form of electromagnetic energy-flux density $|\vec{E}\times\vec{H}|$ which was similar to the electrical power $V(t)\cdot I_R(t)$, the roles played by electric-field and magnetic-field were equivalent to each other in $|\vec{E}\times\vec{H}|$.

## 4 Theoretical Verification of no fringes in standing-waves

**[i]** *The interference fringes described by general formula*

When two monochromatic light-beams $S_1$ and $S_2$ with the electric-field vectors along z-axis were superimposed in their intersection area as shown in Fig.2.1, the total intensity $I$ at P was formulated[9].

$$I = I_1 + I_2 + 2\sqrt{I_1 I_2}\cos\delta \quad (4.1)$$

where $I_1 = |\vec{E_1}|^2$, $I_2 = |\vec{E_2}|^2$ and $\delta$ was the phase difference at the point P between the waves $S_1$ and $S_2$.

In the special case, letting $I_1=I_2$ and setting $\delta$ to zero at the origin (0,0), Eq.4.1 became to

$$I(y) = 2I_1[1+\cos(2\pi\frac{y}{e})] \quad (4.2)$$

where the interference-fringe spacing was

$$e = \frac{\lambda}{2\sin\frac{\theta}{2}} \quad (4.3)$$

From Eq.4.2, we could know that the alternating bright and dark interference-fringes in the intersection area were parallel to x-axis.

The bright fringe's intensity was given by $I(y_{\text{bright}}) = 4I_1$.

The dark fringe's intensity was given by $I(y_{\text{dark}}) = 0$.

As the widths of a bright and a dark fringe were equal to each other, the average of the intensity on yz-plane was given by $2I_1$ [$I_{av}=(4I_1+0)/2$] which was just right consistent with the sum of the intensities of light-beams $S_1$ and $S_2$.

**[ii]** *The inconsistency with the conservation of energy*

When angle $\theta$ became bigger, we could get the intensity average of constant $2I_1$ on yz-plane.

However, the result should be inconsistent with the conservation of energy because the area crossed by the light-beams $S_1$ and $S_2$ on yz-plane became bigger.

**[iii]** *Missing of an inclination factor in common interference formulas*

If the two light-beams $S_1$ and $S_2$ were in the same diameter of $2R$, the energy-flux of each light-beam through yz-plane was also $I_1\cdot(\pi R^2)$.

As $DD = \frac{2R}{\cos(\frac{\theta}{2})}$ where $DD$ was the length of elliptical major axis of each light-beam on yz-plane, the energy-flux-density of each light-beam on yz-plane was $I_{1-yz} = \frac{I_1\cdot(\pi R^2)}{\pi R\cdot(DD/2)} = I_1\cdot\cos(\frac{\theta}{2})$.

Therefore, comparing with Eq.4.2, the interference-fringe's intensity $I$ at any point P on yz-plane should be given by

$$I(y) = 2I_1\cdot\cos(\frac{\theta}{2})[1+\cos(2\pi\frac{y}{e})] \quad (4.4)$$

When the inclination factor $\cos(\theta/2)$ was introduced to the expression in Eq.4.4, the light intensity $I(y)$ could be back to its original meaning of the definition. The light intensity $I$ had been originally defined as 'the time average of the amount of energy which crosses in unit time a unit area perpendicular to the direction of the energy flow'[9].

**[iv]** *No fringes in a standing wave*

From Eq.4.4, $I(y_{\text{dark}}) = 0$;

and $I(y_{\text{bright}}) = 4I_1\cos(\frac{\theta}{2})$.

When $\theta=180°$ that meant a standing-wave field had been formed, $I(y_{\text{bright}}) = 0$ that meant there were no bright-fringes in the standing-wave field.

The result was consisted with zero of the electromagnetic energy-flux-density in a standing-wave field. In other words, there were no alternating bright and dark interference-fringes in a standing wave.

**[v]** *A simple way to understand no fringes in a standing wave*

While angle $\theta$ became bigger in Fig.2.1, the average intensity on yz-plane would be getting weaker because of less illumination (lux) on it according to geometrical optics. As the illumination was in direct proportion to $\cos(\theta/2)$, the intensity on yz-plane would be equal to zero when $\theta=180°$.

**[vi]** *Another way to understand simply*

As known, the alternating bright and dark interference-fringes in the intersection area were parallel to x-axis as shown in Fig.2.1. When the light-beams $S_1$ and $S_2$ traveled into the intersection area, the light-energy would flow along the bright fringes through yz-plane because light-energy would not traverse the dark fringes otherwise the fringes could not keep 'dark'.

While angle $\theta$ became bigger, the bright fringes kept parallel to x-axis. When $\theta=180°$, the light-beams, $S_1$ and $S_2$, were perpendicular to the bright fringes or x-axis. In this case, there was no way to supply light-energy into the bright fringes by the light-beams.

Therefore, it verified that no interference-fringes exist in a standing-wave.

The strict mathematical derivation for the expression in Eq.4.4 was presented in APPENDIX A and APPENDIX C[11]. According to the derivation in the appendixes, alternating bright and dark interference-fringes were significantly formed by interference-energy-flux density $\vec{E} \times \vec{H}$ rather than electric-field energy-density $|E|^2$ alone. The magnetic-field vector acts the same role as the electric-field vector on light interacting with substance.

## 5 The Other Experimental Proofs

As we believed, the ideal experiment in section 2 and the theoretical proof in section 4 were enough to argue that no interference-fringes exist in standing-wave fields, though we still would like to present the other experiments for the verification of the conclusion.

In order to avoid the reflection from the photographic plate, the incident angle on it was set to 57.5° equal to the Brewster's angle of the photographic plate, when it was placed in a pre-adjusted standing-wave field. The thick photographic emulsion was applied instead of the thin film in Wiener's experiment. Thus, instead of the equal thickness fringes of Wiener's experiment, the fringes formed by standing-waves, if they existed really, should be recorded in the thick emulsion. As the fringes in the thick emulsion were equidistant, they would be an equivalent grating similar to a fiber-grating. Then the Bragg-diffraction from the grating could tell us how the grating was.

Because the surface of the emulsion was not parallel to the glass surface on the other side of the Photographic-Plate usually, the Photographic-Plate became a glass wedge actually. The wedge would destruct the pre-adjusted standing-wave field.

In the previous experiments [11], the Photographic Plates had been custom-made to reach better quality of parallelism between the two planes than ordinary Photographic-Plates. 72 photos were taken, in which Bragg-diffraction could not be detected on one spot at least in 11 photos. The equal thickness fringes on the 11 photos illuminated by an expanded light-beam told us that the parallelism between the two sides of the photos was below 4 arc seconds.

It was much more difficult to verify something inexistent than existent. In order to obtain more photos of no-fringes, the idea for the other experiments was come from the performance of interference energy-flux on the surface of a plane mirror. When the incident light-beam illuminated the plane mirror at a small angle, there would be a group of alternating bright and dark fringes in the area of the intersection between the incident and the reflected light-beams. A dark fringe would appear on the surface of the plane mirror if the incident light-beam was in S polarization (electric-field vector was vertical to the incident plane); and a bright fringe would be appear on the surface of the plane mirror if the incident light-beam was in P polarization (magnetic-field vector was vertical to the incident plane), as shown in Fig.5.1.

In this way, the bright fringes formed by light-beam in P polarization were just placed on the dark fringes by light-beam in S polarization. Or the interference-fringes in S and P polarizations overlapped on each other. Thus, if making the photos expose in P and S polarizations alternately, the interference-fringes formed by the incident light-beam in P and S polarizations respectively would cancel each other. Then the more photos of no-fringes could be obtained.

In the experiments, 64 photos were taken aggregately. Bragg-diffraction could not be detected on a spot at least in 17 photos of the 64 photos. The Photographic-Plates used in the experiments were ordinary ones in market rather than custom-made. The equal thickness fringes on the 17 photos showed that the parallelism between the two sides of the photos was below 1 arc minute rather than below 4 arc seconds.

It should be noted that the design concept of the successful experiments was based on the analysis of interference-energy-flux rather than on the establishment of exact standing-wave field. Thus, our idea of the interference-energy-flux density has been verified indirectly.

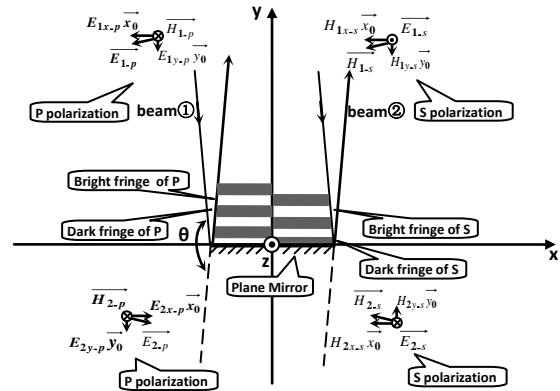

Fig.5.1. The energy-flux in the field of the interference between the incident and reflected beams with P or S polarization

The detail about the experiments was presented in Appendix B.

## 6 Conclusions

With the inspiration of Galileo's ideal experiment of gravity, a novel ideal experiment of standing wave was proposed to significantly present the conception of no-interference-fringes in a standing-wave. The demonstrations have therefore been against our conventional conceptions of interference fringes in standing-waves.

By means of the equivalence between an $LC$ resonance circuit and a pair of antinodes in a standing-wave, we could recognize the characteristics of standing-waves. One of them was that there were no alternating bright and dark fringes in a standing-wave as a form of energy storages. The other one was that electromagnetic energy-flux density was the fundamental-factor to blacken the photographic emulsion rather than electric-field energy-density alone.

The common formula of two-beams-interference was corrected by an inclination factor, so that the meaning of light-intensity was returned to its original definition based on electromagnetic energy-flux density. Missing the inclination factor was due to having comprehensively believed that only electric-field of light blackened the photographic emulsion since 1890 Wiener's experiment. According to the corrected formula of two-beams-interference, we could realize that no bright fringes in a standing wave.

We also demonstrated the other experiments to get higher probability than the previous experiment for obtaining no-Bragg-diffraction photos. In fact, a standing-wave was very difficult to establish in photographic emulsion because the beforehand standing-wave field was often destructed by the glass wedge which was formed by the Photographic-Plate placed in afterward.

In a word, instead of electric-field energy density, interference-energy-flux density is the fundamental-factor in forming interference fringes. The magnetic-field vector acts the same role as the electric-field vector on light interacting with substance.

## 7 A Deduction Plus

The above conclusions seem to trigger the questions on the relevant interpretation in quantum-mechanics, e.g., about the probability density function as follows.

The current interpretation of quantum-mechanics about the probability density function ($|\psi|^2$ in the Schrödinger equation) was formulated by Great Max Born in July 1926[10], who was one of the authors of PRINCIPLES of OPTICS [9].

Apparently, $|\psi|^2$ was educed from $|\vec{E}|^2$ in the classical wave function.

But so far it would not be able to give proper explanation for optical interference including all cases presented in this paper. From our point of view as stated in the verification, the probability density function should be modified based on the energy-flux-density or Poynting vector $|\vec{E} \times \vec{H}|$.

It was for sure that wave-theory was so far imperfect, so that some physical pictures in quantum-mechanics had confused people. For another example, a standing wave was in a laser cavity as known, then there was a series of electric-field nodes. The probability density $|\psi|^2$ at any electric-field node was equal to zero, which meant no photon appeared. A puzzle was initiated for how photons shuttled back and forth through the nodes in a laser cavity?

**Acknowledgment.** Dr. GuanHong Gao made modification of English in this paper. Dr. JiQing Chen proposed beneficial suggestions for modification of this paper and also made modification of English.

REFERENCES
[1] Wiener, "Stehende Lichtwellen und die Schwingungsrichtung polarisirten Lichtes", Annalen der Physik und Chemie 40, 203 (1890).
[2] G. Lippmann, "La photographie des couleurs", C. R. Hebd. Séanc. Acad. Sci. 112, 274 (1891).
[3] P. Drude and W. Nernst, "Ueber die Fluorescenzwirkungen stehender Lichtwellen", Annalen der Physik und Chemie 45, 460 (1892).
[4] H. E. Ives and T. C. Fry, "Standing Light Waves; Repetition of an Experiment by Wiener, Using a Photoelectric Probe Surface", J. Opt .Soc .Amer. 23, 73 (1933).
[5] M. Sasaki, X. Mi and K. Hane, "Standing wave detection and interferometer application using a photodiode thinner than optical wavelength", Appl. Phys. Lett. 75, 2008 (1999).
[6] H. Stiebig, H.-J. Büchner, E. Bunte, V. Mandryka, D. Knipp and G. Jäger, "Standing wave detection by thin transparent n–i–p diodes of amorphous silicon", Thin Solid Films 427, 152 (2003).
[7] H. J. Büchner, H. Stiebig, V. Mandryka, E. Bunte and G. Jäger, "An optical standing-wave interferometer for displacement measurements", Meas. Sci. Technol. 14, 311 (2003).
[8] H. Stiebig, H. Büchner, E. Bunte, V. Mandryka, D. Knipp and G. Jäger, "Standing-wave interferometer", Appl. Phys. Lett. 83, 12 (2003).
[9] M. Born and E. Wolf, PRINCIPLES of OPTICS. (Cambridge University Press, Cambridge, UK, 1999, 7th (Expanded) Ed.), Chap. 7.4, 7.2.
[10] M. Born, "Zur Quantenmechanik der Stoßvorgänge", Zeitschrift für Physik, 37, 863(1926)
[11] JiWu Chen, "The fundamental factor of optical interference" [https://arxiv.org/pdf/1701.08135]

## APPENDIX A: Derivation for The Energy-Flux-Density in Interference Fields and No Energy-Flux in Standing Waves

Let's assume that two parallel linearly polarized light-beams, $S_1$ and $S_2$, in xy-plane with angle $\theta$ between them and with the same wavelength $\lambda$ and with the magnetic-field vectors along z-axis are illuminating yz-plane as shown in Fig.A.1. If the two electric-field vectors of the two beams are denoted as $\vec{E_1}$ and $\vec{E_2}$ respectively and the two magnetic-field vectors as $\vec{H_1}$ and $\vec{H_2}$ respectively, they could be expressed by

$$\vec{E_1} = \vec{E_{10}} \cdot \exp[ik(x\cos\tfrac{\theta}{2} - y\sin\tfrac{\theta}{2})]$$

or

$$\vec{E_1} = \vec{E_{10}} \cdot \cos[\omega t - k(x\cos\tfrac{\theta}{2} - y\sin\tfrac{\theta}{2})] \quad (A1)$$

$$\vec{H_1} = \vec{H_{10}} \cdot \exp[ik(x\cos\tfrac{\theta}{2} - y\sin\tfrac{\theta}{2})]$$

or

$$\vec{H_1} = \vec{H_{10}} \cdot \cos[\omega t - k(x\cos\tfrac{\theta}{2} - y\sin\tfrac{\theta}{2})] \quad (A2)$$

$$\vec{E_2} = \vec{E_{20}} \cdot \exp[ik(x\cos\tfrac{\theta}{2} + y\sin\tfrac{\theta}{2})]$$

or

$$\vec{E_2} = \vec{E_{20}} \cdot \cos[\omega t - k(x\cos\tfrac{\theta}{2} + y\sin\tfrac{\theta}{2})] \quad (A3)$$

$$\vec{H_2} = \vec{H_{20}} \cdot \exp[ik(x\cos\tfrac{\theta}{2} + y\sin\tfrac{\theta}{2})]$$

or

$$\vec{H_2} = \vec{H_{20}} \cdot \cos[\omega t - k(x\cos\tfrac{\theta}{2} + y\sin\tfrac{\theta}{2})] \quad (A4)$$

where, $|\vec{E_{10}}| = |\vec{E_{20}}| = A$, $|\vec{H_{10}}| = |\vec{H_{20}}| = B$, $\omega = 2\pi\nu$, $k = \tfrac{2\pi}{\lambda}$, and $\nu$ is the frequency of the light.

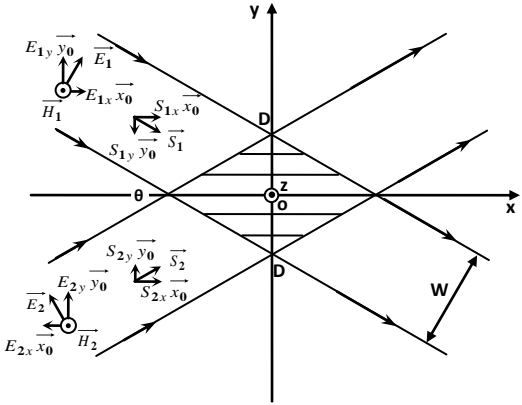

**Fig.A.1. A interference field formed by two parallel light beams in P polarization**

By resolving the energy-flux-densities or Poynting vectors $\vec{S_1}$ and $\vec{S_2}$ into the components, $S_{1x}$ and $S_{2x}$ in the bisector direction of the two beams or along x-axis, and $S_{1y}$ and $S_{2y}$ along y-axis respectively, as shown in Fig.A.1, we can get:

$$\vec{S_1} = S_{1x}\vec{x_0} + S_{1y}\vec{y_0} \quad \text{and} \quad \vec{S_2} = S_{2x}\vec{x_0} + S_{2y}\vec{y_0} \quad (A5)$$

As $\vec{H_1} = H_{1z}\vec{z_0}$ and $\vec{H_2} = H_{2z}\vec{z_0}$, the resolution Eq.A5 of Poynting vectors corresponds to resolving only the electric-field vectors into the components Eq.A6:

$$\vec{E_1} = E_{1x}\vec{x_0} + E_{1y}\vec{y_0} \quad \text{and} \quad \vec{E_2} = E_{2x}\vec{x_0} + E_{2y}\vec{y_0} \quad (A6)$$

where $E_{1x} = A\sin\tfrac{\theta}{2} \cdot \exp[ik(x\cos\tfrac{\theta}{2} - y\sin\tfrac{\theta}{2})]$

and $E_{1y} = A\cos\tfrac{\theta}{2} \cdot \exp[ik(x\cos\tfrac{\theta}{2} - y\sin\tfrac{\theta}{2})]$

and $E_{2x} = -A\sin\tfrac{\theta}{2} \cdot \exp[ik(x\cos\tfrac{\theta}{2} + y\sin\tfrac{\theta}{2})]$

and $E_{2y} = A\cos\tfrac{\theta}{2} \cdot \exp[ik(x\cos\tfrac{\theta}{2} + y\sin\tfrac{\theta}{2})]$

and $H_{1z} = B \cdot \exp[ik(x\cos\tfrac{\theta}{2} - y\sin\tfrac{\theta}{2})]$

and $H_{2z} = B \cdot \exp[ik(x\cos\tfrac{\theta}{2} + y\sin\tfrac{\theta}{2})]$

Then in the intersecting area of the beams, the superimposition of the electric-fields and magnetic-field are (the expressions in complex number space has been converted into real number space directly),

$$E_x = E_{1x} + E_{2x} = 2A\sin\tfrac{\theta}{2} \cdot [-i\sin(ky\sin\tfrac{\theta}{2})]\exp[ikx\cos\tfrac{\theta}{2}]$$

or

$$E_x = -2A\sin\tfrac{\theta}{2}\sin(ky\sin\tfrac{\theta}{2}) \cdot \sin[\omega t - kx\cos\tfrac{\theta}{2}] \quad (A7)$$

and

$$E_y = E_{1y} + E_{2y} = 2A\cos\tfrac{\theta}{2} \cdot [\cos(ky\sin\tfrac{\theta}{2})]\exp[ikx\cos\tfrac{\theta}{2}]$$

or

$$E_y = 2A\cos\tfrac{\theta}{2}\cos(ky\sin\tfrac{\theta}{2}) \cdot \cos[\omega t - kx\cos\tfrac{\theta}{2}] \quad (A8)$$

and

$$H_z = H_{1z} + H_{2z} = 2B \cdot [\cos(ky\sin\tfrac{\theta}{2})]\exp[ikx\cos\tfrac{\theta}{2}]$$

or

$$H_z = 2B\cos(ky\sin\tfrac{\theta}{2}) \cdot \cos[\omega t - kx\cos\tfrac{\theta}{2}] \quad (A9)$$

Thus, the instantaneous energy flux density along y-axis is

$$S_y = E_x \cdot H_z = -AB\sin\tfrac{\theta}{2}\sin(2ky\sin\tfrac{\theta}{2}) \cdot \sin[2(\omega t - kx\cos\tfrac{\theta}{2})] \quad (A10)$$

Eq.A10 indicates the presence of the standing-wave component in the interference field or the intersecting area of the beams, as the electric-field component $E_x$ is in quadrature with the magnetic-field vector $H_z$. Then its time-averaged value of the energy-flux-density is

$$\langle S_y \rangle_T = 0 \quad (A11)$$

Eq.A11 is consistent with the summation of ($S_{1y} + S_{2y}$) along y-axis which is $S_y = S_{1y} + S_{2y} = 0$ as shown in Fig.A.1.

The corresponding instantaneous energy-flux-density through yz-plane along x-axis, called Interference-Energy-Flux-Density, is, $S_x = E_y \cdot H_z$, then

$$S_x = 4AB\cos\tfrac{\theta}{2}\cos^2(ky\sin\tfrac{\theta}{2}) \cdot \cos^2(\omega t - kx\cos\tfrac{\theta}{2}) \quad (A12)$$

where the electric-field component $E_y$ is in phase with the magnetic-field vector $H_z$. The time-averaged value of the energy-flux-density along x-axis is,

$$\langle S_x \rangle_T = 2AB\cos\tfrac{\theta}{2}\cos^2(ky\sin\tfrac{\theta}{2}) \quad (A13)$$

or

$$\langle S_x \rangle_T = AB\cos\tfrac{\theta}{2}[1 + \cos(2ky\sin\tfrac{\theta}{2})] \quad (A14)$$

or

$$\langle S_x \rangle_T = \sqrt{\tfrac{\varepsilon}{\mu}}A^2\cos\tfrac{\theta}{2}[1 + \cos(2ky\sin\tfrac{\theta}{2})] \quad (A15)$$

$S_x = S_{1x} + S_{2x}$ is the interference energy-flux-density according to Fig.A.1. The interference result of the two parallel light beams is shown by Eq.A13~A15. It has missed the ratio of $\cos(\theta/2)$ in the general equation derived only from the electric-fields. Eq.A13~A15 are the distribution of the energy-flux-density in the interference field including electric and magnetic fields rather than electric-field energy density alone.

Therefore, in the intersecting area of the two beams, it includes two components, traveling-wave and standing-wave.

The traveling-wave is interference energy-flux-density $S_x = E_y \cdot H_z$, which makes interference fringes as well known. The average of the interference energy-flux-density is presented by Eq.A13~A15, in which appeared an aslant coefficient of $\cos(\theta/2)$, the term more than the general formula of two-beams-interference.

The standing-wave is $S_y = E_x \cdot H_z$ with zero average of energy-flux-density as shown by Eq.A11. The energy-density of electric-field in the standing-wave is:

$$\frac{\varepsilon}{2}|E_x|^2 = \frac{\varepsilon}{2}\left[2A\sin\frac{\theta}{2}\sin(ky\sin\frac{\theta}{2}) \cdot \sin(\omega t - kx\cos\frac{\theta}{2})\right]^2 \quad (A16)$$

And because the energy-density of the magnetic-field is equal to the energy-density of the electric-field in the standing-wave, the energy-density of the magnetic-field is:

$$\frac{\mu}{2}|H_{z-SW}|^2 = \frac{\mu}{2}\left[2B\sin\frac{\theta}{2}\cos(ky\sin\frac{\theta}{2}) \cdot \cos(\omega t - kx\cos\frac{\theta}{2})\right]^2 \quad (A17)$$

Compared by Eq.A13 and Eq.A17, it can be observed that the interference energy-flux-density $S_x = E_y \cdot H_z$ is traveling just along the antinodes of the magnetic-field in the standing-wave component.

When $\theta = 180°$ that means the two beams has turned into a standing-wave, $S_x \equiv 0$ at any time according to Eq.A12 or the interference energy-flux-density is equal to zero. That is, there are no bright and dark fringes in a standing-wave.

According to Eq.A12 and Eq.A13, it has been realized that interference-energy-flux density is the fundamental-factor in forming interference fringes rather than electric-field energy density alone. The magnetic-field vector acts the same role as the electric-field vector on light interacting with substance.

## APPENDIX B: The Other Experiments

### B.1 The Difficulty For Photos Exposed In an Exact Standing-wave

In order to prove experimentally that no standing-wave's fringes existed in the emulsion, the standing-wave's field had to be set up before a Photographic-Plate was placed in the field for exposing to record the fringes by the standing-wave. However, because the surface of the emulsion was not parallel to the glass surface on the other side of the Photographic-Plate normally, the Photographic-Plate became a glass wedge actually. So the wedge would deflect the incident beam① and the reflected beam②, as shown in Fig.B.1. In other words, the wedge had broken the standing-wave field which was pre-adjusted under the condition that the reflected beam② was just traveling against the input beam①. In this way, the Photographic-Plates had often recorded general interference fringes instead of standing-wave's fringes.

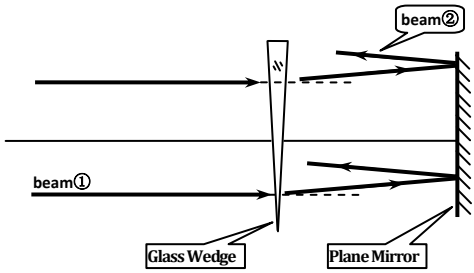

**Fig.B.1 The direction of beam② was deflected from its pre-adjusted path as the Photographic-Plate became a glass wedge**

### B.2 The Bright Fringes' Courses Depending on The Polarization of Two Parallel Light-beams

#### (1) S polarization

If two parallel linearly-polarized light-beams $\vec{S_1}$ and $\vec{S_2}$ in xy-plane with the angle $\theta$ between them and with the same wavelength $\lambda$ and with the electric-field vectors ($\vec{E_1}$ and $\vec{E_2}$) along z-axis were illuminating yz-plane, a group of alternating bright and dark interference-fringes would appear in the intersecting area, as shown in Fig.B.2.

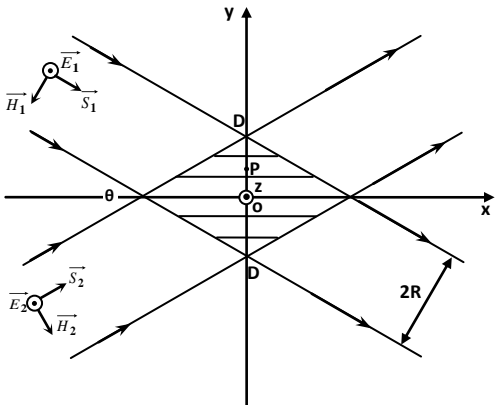

**Fig.B.2. An interference field formed by two parallel light beams in S polarization**

The vibrating state of electric-field vector perpendicular to the incidence plane was called as S polarization generally.

In the intersecting area, the electric-field and the components of the energy-flux density could be expressed as [APPENDIX C in 11]

$$E_z = 2a\cos(ky\sin\frac{\theta}{2})\cdot\cos(\omega t - kx\cos\frac{\theta}{2}) \quad (B1)$$

$$S_x = 4ab\cos\frac{\theta}{2}\cos^2(ky\sin\frac{\theta}{2})\cdot\cos^2(\omega t - kx\cos\frac{\theta}{2}) \quad (B2)$$

$$S_y = ab\sin\frac{\theta}{2}\sin(2ky\sin\frac{\theta}{2})\cdot\sin[2(\omega t - kx\cos\frac{\theta}{2})] \quad (B3)$$

where $a$ and $b$ were the amplitudes of the electric-field and the magnetic-field respectively and $k = \frac{2\pi}{\lambda}$.

From Eq.B3, the time-averaged value of the energy-flux-density along y-axis could be

$$\langle S_y \rangle_T = 0 \quad (B4)$$

It meant there was a standing-wave as an energy-flux-density component along y-axis in the intersecting area.

From Eq.B2, the time-averaged value of the energy-flux-density along x-axis was

$$\langle S_x \rangle_T = ab\cos\frac{\theta}{2}[1+\cos(2ky\sin\frac{\theta}{2})] \quad (B5)$$

When $2ky\sin\frac{\theta}{2} = (2m+1)\pi$ where $m$ was integer, we could get $\langle S_x \rangle_T = 0$ which meant the dark fringe courses parallel to x-axis and at the nodes of the electric-field [$E_z$(at a node) = 0] along y-axis.

When $2ky\sin\frac{\theta}{2} = 2m\pi$, $\langle S_x \rangle_T$ = maximum which meant the bright fringe courses parallel to x-axis and at the antinodes of the electric-field ($|E_z|$ reached its maximum) along y-axis.

#### (2) P polarization

The vibrating state of electric-field vector parallel to the incidence plane was called as P polarization generally. When the electric-field vector and the magnetic-field vector were interchanged in Fig.B.2, the interference field was formed by the two parallel light-beams as shown in Fig.A.1 [APPENDIX A]. In the intersecting area, the magnetic-field and components of the energy-flux density could be expressed as [APPENDIX A]

$$H_z = 2B\cos(ky\sin\frac{\theta}{2})\cdot\cos(\omega t - kx\cos\frac{\theta}{2}) \quad (B6)$$

$$S_x = 4AB\cos\frac{\theta}{2}\cos^2(ky\sin\frac{\theta}{2})\cdot\cos^2(\omega t - kx\cos\frac{\theta}{2}) \quad (B7)$$

$$S_y = AB\sin\frac{\theta}{2}\sin(2ky\sin\frac{\theta}{2})\cdot\sin[2(\omega t - kx\cos\frac{\theta}{2})] \quad (B8)$$

where $A$ and $B$ were the amplitudes of the electric-field and the magnetic-field respectively.

Similarly, $\langle S_x \rangle_T = 0$ indicated the dark fringe courses parallel to x-axis and at the nodes of the magnetic-field [$H_z$(at a node) = 0] along y-axis.

$\langle S_x \rangle_T$ = maximum indicated the bright fringe courses parallel to x-axis and at the antinodes of the magnetic-field ($|H_z|$ reached its maximum) along y-axis.

Therefore, bright fringes' courses depended on polarization of two parallel light-beams forming interference field.

### B.3 Experimental Design

The objective of the experiments was to record fringes of standing-wave in thick photographic emulsion and to show no fringes in the emulsion. If there were interference fringes in thick photographic emulsion, the equidistant fringes would be an equivalent grating similar to a fiber-grating. Then the Bragg-diffraction from the grating could tell us how the grating was. Specially, if no Bragg-diffraction was detected, we could affirm that no fringes were in the emulsion of the photo.

However, an exact standing-wave field was very difficult to establish as the Photographic-Plate would become an equivalent glass-wedge after it was placed in the optical system for exposing.

The experiments were based on different courses of the bright fringes by two light-beams changing with S or P polarization as discussed in section B.2. While the incident and the reflected beams were superimposed onto each other at a plane mirror, the boundary condition on the mirror was that the electric-field sum component parallel to the mirror surface would be equal to zero. The boundary warranted that a dark fringe (zero of energy-flux) would appear on the mirror surface when the incident light-beam was in S-polarization and a bright fringe (maximum of energy-flux) would flow along the surface in P-polarization, as shown in Fig.B.3.

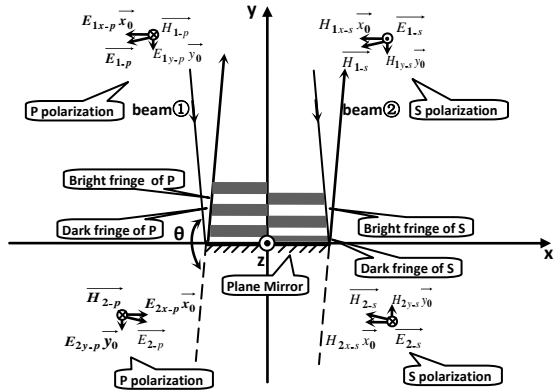

**Fig.B.3. The energy-flux in the field of the interference between the incident and reflected beams with P or S polarization**

The spacing of the fringes in the emulsion would almost be equal to $\lambda/2$ of the spacing in the standing-wave field, as the deflected angle brought by the equivalent glass-wedge was less than 1 arc minute.

If a light-beam ① with S polarization illuminated a plane mirror at a small angle, the interference fringes would be formed by the incident beam ① and the reflected beam ②, as shown in the right side of Fig.B.3. If a light-beam ① with P polarization illuminated a plane mirror at a small angle, the interference fringes would be formed by the incident and the reflected beams as shown in the left side of Fig.B.3. The two kinds of fringes were just interlaced because the magnetic-field antinodes were just at the electric-field nodes (the first electric-field node was on the Plane-Mirror). If a Photographic-Plate in the interference field was exposed in the light-beams with S-polarization and in P-polarization respectively, the two kinds of fringes would cancel each other.

In this way, we would obtain more photos without Bragg-diffraction than the previous experiments[11].

### B.4 Setup

The arrangement for the experiments was illustrated in Fig.B.4. The Laser Beam from Newport Model: R-30991 (HeNe Laser, 633 nm, 5.0 mW, 500:1 Polarization, Longitudinal Mode: 441 MHz) was split by the Polarizing Beam Splitter (PBS) into two beams. The one beam through the PBS was passing through the P Polarizer for much purer P-polarization in the plane of incidence, and the other one reflected by the PBS was passing through the S Polarizer. Then each beam, passing through or reflected from the Half Reflecting Mirror, was focused by the Microscope Objective Lens on the Pinhole with a diameter of 5μm. And they were subsequently expanded by Telescope Objective Lens1 with a focal length of 382mm and a diameter of 65mm to form a parallel beam with a diameter of 65mm. The parallel beam with the Aperture2 of 25mm, through the Photographic-Plate, normally illuminated the Plane-Mirror surface. The reflected light from the Plane-Mirror traveled back in its coming way and superimposed onto the incident light beam to form an optical standing-wave in the Photographic Emulsion which could show the profile of the standing wave after development.

1870mm and 680mm were set for ensuring coherence between the incident and reflected light beams in the Emulsion. The incidence angle β was set to the Brewster's angle of 57.5° for avoiding the reflection by the surfaces of the Photographic-Plate.[11]

The reflected light from the Plane-Mirror would be back focused on the Pinhole by Telescope Objective Lens1, and the focused light point was actually the image of the Pinhole. The reflected light of the image by the surface of the Pinhole would interfere with the light directly from the Pinhole, and their interference fringes could be seen on a screen close to Telescope Objective Lens1 (the screen was not drawn in Fig.B.4). For assuring that the normal of the Plane-Mirror surface was parallel to the optical axis as exactly as possible, the Plane-Mirror should be adjusted slowly, so that the interference fringes were as few as possible and the direction of the fringes was kept level, which meant the normal of the Plane-Mirror was inclined at a small angle from level in the sagittal plane of Fig.B.4.

### B.5 Results of The Photos Without Bragg-Diffraction

**The experiments and results**

Before taking photos, the 45⁰ Polarizer in Fig.B.4 should be adjusted to make the P beam (through P Polarizer) and the S beam (through S Polarizer) equal to each other in light intensity. The exposure time was controlled by an electric shutter in order to ensure the amount of exposure in the linearity region of the emulsion. Eight photographs were exposed on every Photographic-Plate with a rectangular shape of 90mm×240mm at the incidence angle of 57.5°. Among them, the four photographs were exposed to the light-beams with P and S polarization alternately in the same amount of exposure by switching Black-Paper on their optical path. This meant that each of the four photographs was exposed in a half exposure by the light beam in P polarization while S Black Paper was blocking the S polarization path, and then alternately exposed in the other half exposure by the light in S polarization. The four photographs would be denoted as PS. As comparison, the other four photographs were divided into two groups which were exposed only in P or S polarization respectively and were denoted as PP or SS. In P or S polarization only, the photographs also were exposed twice in half exposure time.

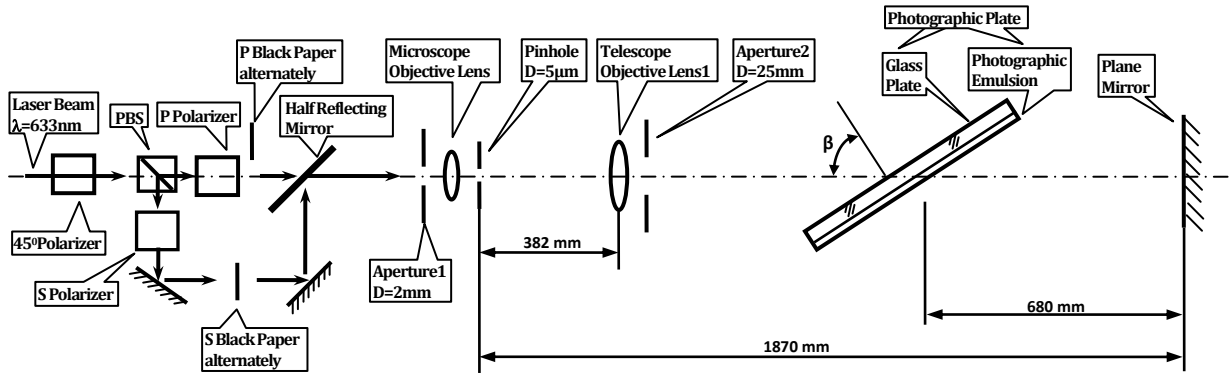

**Fig.B.4. The setup for experiments on exact optical standing waves (mirrors used to fold the optical path were not illustrated)**

In order to verify if the interference fringes equivalent to a Bragg grating by the standing-wave existed in the emulsion, the Bragg-diffraction of every photo was examined by placing the developed Photographic-Plate in the same setup. A paper covered the Plane-Mirror in case of reflection from it. The Microscope Objective Lens and the Pinhole were removed out of the optical path in Fig.B.4 to make only a thin laser beam in P polarization project on the photos. While the laser beam was illuminating the Bragg grating along the optical axis, it would cause the Bragg-diffraction to return back in the coming way of the incident-light. When the developed Photographic-Plate was slightly turned, the Bragg-diffraction, the returning light-beam, could be found near the axis. Only the points at the elliptical major axis on the photos were inspected for convenience.

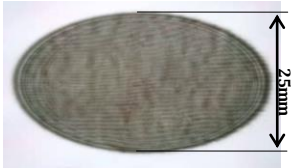 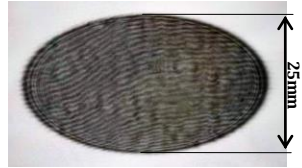

**Fig.B.5. A No-Bragg-diffraction photo in PS**  **Fig.B.6. A No-Bragg-diffraction photo in SS.**

Interference fringes of equal thickness on every photo indicated the angle and direction of the glass-wedge which was formed by the surface of the emulsion and the glass surface on the other side of the Photographic-Plate. The fringes of equal thickness on SS or PS or SP photos were shown on their emulsion side, because those fringes were formed by S-polarization light-beams which could be reflected from both sides of the photo, as shown in Fig.B.5 and Fig.B.6. The contrast-ratio of fringes in Fig.B.5 was lower than in Fig.B.6. The reason was that the exposure only in S-polarization light-beam was able to make the fringes of equal thickness in Fig.B.5 and the exposure in the P-polarization could not. The double exposure in S-polarization light-beam made the higher contrast-ratio of fringes in Fig.B.6 than Fig.B.5.

There were no fringes of equal thickness on any PP photos because no light beams reflected from the both sides of the Photographic-Plate while the incident angle was the Brewster's angle of 57.5°. In order to check the fringes of equal thickness on PP photos, an expanded parallel light beam illuminated on every PP photo. And then the fringes of equal thickness could be shown on a screen where the reflected light beams from both sides of the photos were projected.

**Table 1. Statistics of the results with no and weak Bragg-diffraction from the photos exposed in the ways of PS or SP and PP or SS**

| Exposed way | PS or SP | SS | PP |
|---|---|---|---|
| No Bragg-diffraction | 7 | 9 | 1 |
| Weak Bragg-diffraction (light power less than 4nW) | 12 | 3 | 3 |
| Number of total photos | 32 | 16 | 16 |

Under illumination of the thin laser beam with a power of about 1mW, the photos with no or weak Bragg-diffraction were enumerated in Table 1. "No Bragg-diffraction" meant that there was at least one spot without Bragg-diffraction on the photo. "Weak Bragg-diffraction" meant that the light power of the Bragg-diffraction at a spot of the photo was less than 4nW.

**The reason for no-Bragg-diffraction**
The normal of the Plane-Mirror had been adjusted in the sagittal plane of Fig.B.4 as mentioned in the section B.4 before a Photographic-Plate was placed in the experimental system.
*(1) Exposed in PS or SP*
Among 32 photos exposed in PS or SP way as shown in Table 1, the spots without Bragg-diffraction were detected on 7 photos and the spots with weak-Bragg-diffraction were on 12 photos. On the most of (7+12) photos, the directions of their equal-thickness fringes were almost horizontal as shown in Fig.B.5. That indicated the ridge of the photo's glass-wedge was perpendicular to the plane of Fig.B.1, or the normal of the Plane-Mirror was in the sagittal plane of Fig.B.4. Thus, the interference fringes formed in P and S polarizations alternatively were similar to Fig.B.3 and the spots with no or weak Bragg-diffraction existed in the photos with high probability.
*(2) Exposed in SS*
As shown in Table 1, among 16 photos exposed in SS way, the spots without Bragg-diffraction were detected on 9 photos, and the spots with weak-Bragg-diffraction were on 3 photos. On the most of (9+3) photos, the directions of their equal-thickness fringes were inclined at near 45° from the level as shown in Fig.B.6, which indicated the ridge of the glass-wedge was parallel to equal-thickness fringes. If the electric-field vectors in S-polarization were decomposed in the two directions between vertical to and along the ridge, the circumstances in '(1) Exposed in PS or SP' would appear equivalently. The difference was that the photos in SS way were exposed in P and S polarization simultaneously rather than alternatively. Thus, there were no or weak interference fringes in the photos.
*(3) Exposed in PP*
Among 16 photos exposed in PP way, only 4 photos with no or weak Bragg-diffraction were enumerated in Table 1, and all of them had almost vertical fringes of equal-thickness which meant the ridge of the photo's glass-wedge parallel to the plane of Fig.B.1, or vertical to the plane of Fig.B.4.
As the normal of the Plane-Mirror had been pre-adjusted in the sagittal plane of Fig.B.4 additionally, a glass-wedge's ridge was formed with 45° from the level equivalently. It was equivalent in the circumstances in '(2) Exposed in SS', which explained no or weak interference fringes in the photos.
*(4) Two perfect photos without Bragg-diffraction anywhere*
Among 17 photos of no-Bragg-diffraction, there were two perfect photos on which Bragg-diffraction had not been detected at any spot.
One of them had a group of equal-thickness fringes with a rare density of fringe-spacing which meant the glass-wedge angle of the Photographic-Plates was below 4 arc seconds[11]. It indicated that the glass-wedge made the light-beam through the photo illuminate just perpendicularly on the Plane-Mirror and warrant an exact standing-wave field establishing in the emulsion of the photo.
The other one had a very uniform distribution of equal-thickness fringes, which indicated that the glass-wedge just cancelled the pre-adjusted Plane-Mirror's tilt and also made the light-beam through the photo illuminate just perpendicularly on the Plane-Mirror.

In the experiments demonstrated above, we have obtained more photos without Bragg-diffraction than the previous experiments[11]. It should be noted that the design concept of the successful experiments was based on the analysis of interference-energy-flux rather than on the establishment of exact standing-wave field. Thus, our idea of the interference-energy-flux density has been verified indirectly.